\documentclass{pasj02}

\Received{\today}
\Accepted{}
 
\usepackage{threeparttable,booktabs,color,amsmath,url,natbib}

\jyear{2024}
\Received{2024/12/18}
\Accepted{}

\begin{document}

\title{Revisiting the Suzaku spectrum of the Galactic SNR W\,49\,B: non-detection of iron K-shell charge exchange emission and refined ejecta mass ratios of iron-group elements}

\author{Makoto \textsc{Sawada}\altaffilmark{1, 2}%
}
\altaffiltext{1}{Department of Physics, Rikkyo University, 3-34-1, Nishi-ikebukuro, Toshima-ku, Tokyo 171-8501, Japan}
\altaffiltext{2}{Center for Pioneering Research, RIKEN, 2-1 Hirosawa, Wako, Saitama 351-0198, Japan}
\email{makoto.sawada@rikkyo.ac.jp}

\author{Toshiki \textsc{Sato}\altaffilmark{3}}

\altaffiltext{3}{Department of Physics, School of Science and Technology, Meiji University, 1-1-1 Higashi Mita, Tama-ku, Kawasaki, Kanagawa
214-8571, Japan}

\author{Keiichi \textsc{Maeda}\altaffilmark{4}}
\altaffiltext{4}{Department of Astronomy, Kyoto University, Kitashirakawa-Oiwake-cho, Sakyo-ku, Kyoto 606-8502, Japan}

\author{Koki \textsc{Itonaga}\altaffilmark{1}}


\maketitle

\begin{abstract}
The origin of the recombining plasma in several Galactic SNRs has been debated. A plausible mechanism would be a rapid cooling in the past, either by adiabatic or conductive process. A recent spectral study of W\,49\,B reported a possible charge exchange (CX) emission due to collisions between the shock-heated ejecta and external cold clouds, which could be a direct support for the conduction cooling scenario. However, a potentially large systematic uncertainty in the spectral analysis has not been examined. In this paper, we revisit the Suzaku spectrum of W\,49\,B with taking into account the systematic uncertainties in spectral codes and instrumental gain calibration. We find that the previously reported CX flux is fully attributable to dielectronic recombination satellite lines of high-shell transitions that are missing from the present version of the spectral codes. We also report refined Fe-group ejecta mass ratios, which, in comparison to those in the literatures, show a better agreement with theoretical expectations from nucleosynthesis models, either of Type Ia explosions or spherical core-collapse explosions.
\end{abstract}


\section{Introduction} \label{sec:intro}

The dynamical and thermal evolution of supernova remnants (SNRs) has been studied with spectral analysis of shock-heated plasma using a non-equilibrium ionization (NEI) model. A series of Suzaku observations has found over-ionized/recombining plasma from several middle-aged SNRs \citep[e.g.,][]{Yamaguchi2009, Ozawa2009, Ohnishi2011, Sawada2012, Uchida2012}, which was unexpected based on the conventional evolution model of an SNR in a uniform interstellar medium. Since the discovery, the origin of the recombining plasma in SNRs has been debated. The observed charge-state distribution (CSD) higher than that at the collisional ionization equilibrium (CIE) at the observed electron temperature requires a rapid electron cooling or additional ionization source in the past. Recently claimed proton ionization contribution by the low-energy tail of shock-accelerated non-thermal protons \citep{Hirayama2019, Yamauchi2021} has been proved to be not efficient enough to make the observed discrepancy from the CIE charge state \citep{Sawada2024}. Together with the absence of other types of strong ionizing sources (e.g., a bright photo-ionization source) in the vicinity of the recombining SNRs, a rapid electron cooling would be the only remaining possibility. There are two major scenarios employing different physical processes. One is the rarefaction model, in which a rapid cooling of the ions and electrons is caused by the shock interaction with dense circumstellar matter (CSM) made of stellar winds from a massive progenitor star \citep{Itoh1989, Shimizu2012}. The other is the conduction model, in which a collision with cold atomic/molecular clouds is assumed and the thermal conduction plays the role to cool the electrons in shock-heated plasma \citep{Kawasaki2002, Zhou2011}. To date, there has been no decisive conclusion on which scenario is the primary mechanism responsible for the recombining plasma in SNRs.

W\,49\,B is the youngest recombining SNR, whose dynamical age is estimated to be a few~kyr \citep{Hwang2000, Zhou2018}. It has the highest electron temperature ($kT_{\rm e} \approx 1.5$~keV and exhibits strong Fe K-shell emission such as Fe\emissiontype{XXV} He$\alpha$, Fe\emissiontype{XXVI} Ly$\alpha$, and Fe K-shell radiative recombination continuum (RRC), originating from shock-heated ejecta \citep{Ozawa2009}. The origin of the recombining plasma in this remnant has been discussed mainly based on spatial correlations. \citet{Yamaguchi2018} analyzed spatially resolved NuSTAR spectra and showed a clear positive correlation between the electron temperature and recombination timescale, which was interpreted as the evidence for the rarefaction scenario; a volume that rarefied more has a lower density, which is reflected in a lower recombination timescale, and due to the adiabatic cooling, it has a lower temperature. On the other hand, \citet{Sano2021} used the same NuSTAR data but reached to a different conclusion. A region with lower electron temperature was found to have a higher column density of ambient cold clouds detected by ALMA, which was argued to be the evidence for the conduction cooling. A further support for the conduction scenario has been reported most recently by \citet{Suzuki2024}, who argued significant detections of Fe\emissiontype{XXV} charge exchange (CX) emission in 8--9~keV and Fe\emissiontype{I} K$\alpha$ fluorescence line at 6.4~keV from the Suzaku spectrum. The existence of CX emission, in particular, would be a clear indication of the physical interaction between shock-heated ejecta and ambient cold clouds. However, no examination on a possible systematic error in the observed fluxes of these features was conducted, which poses a question on the robustness of the detections, especially for the relatively new spectral feature in the X-ray astronomy, the CX emission.

Another important clue to determine the dominant process for producing the recombining plasma would be the progenitor of the remnant. W\,49\,B was once claimed as a hypernova remnant based on the interpretation of the bar-like morphology found by Chandra observations \citep{Lopez2013} as a result of a jet-driven explosion. On the other hand, recent measurements have shown that the observed elemental pattern is more in line with a Type Ia origin. For instance, \citet{Zhou2018} analyzed Chandra spectra and concluded that the observed abundance pattern, especially a high Mn/Cr ratio, cannot be explained by any core-collapse (CC) models, including bipolar CC models. The discriminating power of the explosion types would be enhanced by combining multiple ratios. \citet{Sato2020} combined two ratios, Mn/Cr and Ni/Fe, to reveal the synthesis site of a Fe-rich structure in the Kepler's SNR. They extended their discussion to other (possible) Ia SNRs, 3C\,397 and W\,49\,B, and pointed out the peculiarity of W\,49\,B based on the Mn/Cr and Ni/Fe ratios complied from the literatures. In Type Ia nucleosynthesis models, the Mn/Cr ratio is often positively correlated with the Ni/Fe ratio, and thus for W\,49\,B which exhibits a high Mn/Cr ratio, the Ni/Fe ratio is also expected to be high. However, the observed value is significantly lower compared to the trend in the model prediction. This means that, although the abundance pattern of a certain range of elements indicates a Type Ia origin, that of the Fe-group elements cannot be reproduced by any of existing Ia models. Although it is possible that the observed, puzzling abundance pattern of the Fe-group elements is indeed the outcome of a unique nucleosynthesis due to a peculiar explosion, another possibility is again the systematics in the spectral analysis; it is possible that those in the spectral model or in the calibration have considerably affected the derived abundance pattern. This would be particularly important for the Ni/Fe ratio because of the blending of Ni\emissiontype{XXVII} He$\alpha$ and Fe\emissiontype{XXV} He$\beta$ with the moderate spectral resolution of X-ray CCDs, which forces us to heavily rely on the spectral models to get the ratio. 

In the recent progress in the X-ray spectroscopy of collisional hot plasmas, it has been emphasized that comparing results with different spectral codes is imperative to obtain a reliable conclusion as demonstrated by \citet{Hitomi2018}. Although the systematic investigation using the Hitomi data has significantly raised the level of the agreement between independently developed spectral codes and the overall accuracy, this improvement is limited to a CIE plasma. As pointed out by \citet{Sawada2019}, the potential systematic uncertainty seems to be still large in the case of an NEI plasma even at the bright line complex such as Fe\emissiontype{XXV} He$\alpha$. In this paper, we revisit the X-ray spectrum of W\,49\,B using the archival Suzaku X-ray Imaging Spectrometer (XIS) data and report spectral analysis including comparison between results with different spectral codes.
This paper is structured as follows. The observations and data reduction are described in \S\ref{sec:obs}. The analysis and results are presented in \S\ref{sec:ana}. The implication of the results will be discussed in \S\ref{sec:disc}, and finally the paper will be summarized in \S\ref{sec:summary}.

\section{Observations and data reduction} \label{sec:obs}

W\,49\,B was observed in 2009 (IDs 503084010 and 504035010 with the total exposure of 113.5~ks) and 2015 (IDs 509001010, 509001020, 509001030, and 509001040 with the total exposure of 400.8~ks). In this paper, we concentrate on the two front-illuminated CCDs (XIS\,0 and XIS\,3) because of their higher quantum efficiencies and lower particle background levels in the Fe K-shell band compared to the back-illuminated one (XIS\,1). Also, we use the data from observations in 2009 only. The choice is because our objective is to revisit the Suzaku spectrum with deeper attention to systematic uncertainties, and thus the priority is to have superior spectral resolution rather than to maximize the statistics. The difference in the energy resolution between the two groups of the observations is indeed large. For instance, for XIS\,3, the resolution at the full width half maximum was measured at $\approx 160$~eV (2009) vs. $\approx 190$~eV (2015) at 5.9~keV using the onboard radioactive calibration sources $^{55}$Fe. This choice degrades the statistical precision. However, as reported by \citet{Suzuki2024}, the data of XIS\,0 from observations in 2015 is not usable, and thus, the degradation factor for the statistical errors is only 1.7, which is, for our purpose, acceptable compromise to have superior energy resolution. 

The retrieved data products have been processed by the latest version (3.0) of the Suzaku pipeline analysis and applied the latest calibration. We started our analysis from cleaned event lists, which have already been filtered using the standard screening criteria. 

\section{Analysis and results} \label{sec:ana}

\subsection{Extraction and background subtraction}

\begin{figure}[htbp]
\begin{center}
\includegraphics[width=8cm]{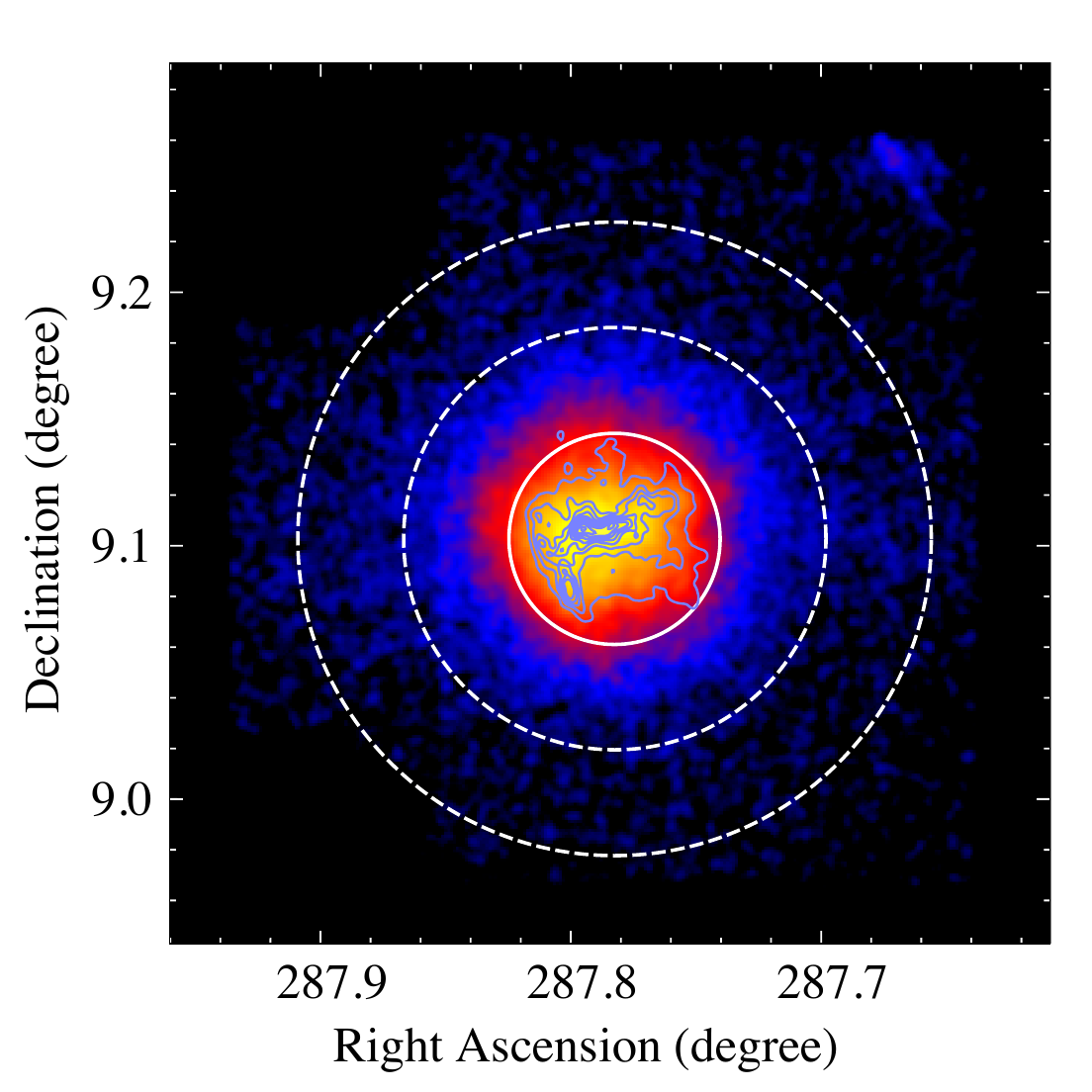}
\end{center}
\caption{X-ray image in 5.0--9.5~keV with XIS\,0 and XIS\,3 combined. The blue contours are from the 4.1--8.0~keV image with Chandra. The source and background regions for spectral analysis are shown with a solid circle and dashed annulus, respectively.
}
\label{fig:image}
\end{figure}

Figure~\ref{fig:image} shows an X-ray image in 5.0--9.5~keV using XIS\,0 and XIS\,3. We extracted the source spectrum from a circular region with a radius of \timeform{2.5'} and the background spectrum from an annulus with an inner and outer radii of \timeform{5.0'} and \timeform{7.5'}, respectively. It is known that the drop in the mirrors' effective area toward higher X-ray energy is severer for larger off-axis angles due to the vignetting effect. Therefore, we corrected this effect in the background subtraction using the method as performed by \citet{Hyodo2008}. More specifically, we generated and subtracted non-X-ray background (NXB) spectra from the source and background spectra using \texttt{xisnxbgen} \citep{Tawa2008} as these are not affected by the vignetting, and then multiplied the source-to-background effective area ratio to the NXB-subtracted spectrum of the background region. Here, to get the effective area ratio, we used auxiliary response files (ARF) for the two regions using \texttt{xissimarfgen} \citep{Ishisaki2007} assuming a sky with a uniform surface brightness. This way, we obtained the background spectrum as if it is observed at the same off-axis angle distribution as the source region and therefore the same vignetting effect. Finally, the vignetting-corrected background spectrum was subtracted from the source spectrum to obtain the background-subtracted spectrum of W\,49\,B. We confirmed that the product is essentially the same as the one analyzed by \citet{Ozawa2009}. 

We analyzed the background subtracted XIS spectrum in 5.0--11.0~keV using a recombining plasma model affected by interstellar absorption. For the analysis, we generated redistribution matrix file (RMF) with \texttt{xisrmfgen} and ARF using the Chandra/ACIS 4.1--8.0~keV image with \texttt{xissimarfgen}. We adopt the proto-solar abundances by \citet{Lodders2009} as the reference solar abundances, the distance to W\,49\,B of 11.3~kpc \citep{Brogan2001, Sano2021}, and the atomic-hydrogen-equivalent interstellar absorption column density of $5\times10^{22}$~cm$^{-2}$ \citep{Keohane2007}. We will quote the statistical errors at the 90\% confidence interval. The recombining plasma model has the initial temperature $kT_{\rm init}$ to set up the initial CSD, current electron temperature $kT_{\rm e}$ to determine the spectral shape contributed by individual ion species, and recombination timescale $\tau$ to obtain the current CSD assuming thermal relaxation under a constant temperature of $kT_{\rm e}$. Other parameters are the chemical abundances $Z_X$ for element $X$ and emission measure $EM$. We allowed abundances of Cr, Mn, Fe, and Ni to be free, while those of other elements were fixed at the proto-solar values. The emission measure was also a free parameter and represented in the form of the volume emission measure assuming the distance to W\,49\,B \citep[11.3~kpc:][]{Brogan2001, Sano2021}. To address possible systematic error in the XIS gain-scale calibration, an adjustment was introduced as a linear stretch in the energy grid of emission models, i.e., $E = (1 - f_{\rm gain}) E_0$, where $E_0$ and $E$ are the energy grids of emission models before and after the adjustment, respectively. Note the negative sign before $f_{\rm gain}$, meaning that a positive $f_{\rm gain}$ indicates underestimation of the detector gain in the calibration. The required adjustment was found to be $f_{\rm gain} = 6.65 \times 10^{-4}$, which is consistent with the relative centroid shift by a factor of $\approx 0.1$\% at 5.9 keV measured with the calibration source $^{55}$Fe. We repeated the analysis with and without this gain adjustment to gauge the magnitude of the systematic uncertainty in the derived spectral parameters due to a possible gain calibration error. 

\subsection{Spectral fits using AtomDB and SPEX} \label{sec:fit}

\begin{figure*}[htbp]
\begin{center}
\includegraphics[width=17cm]{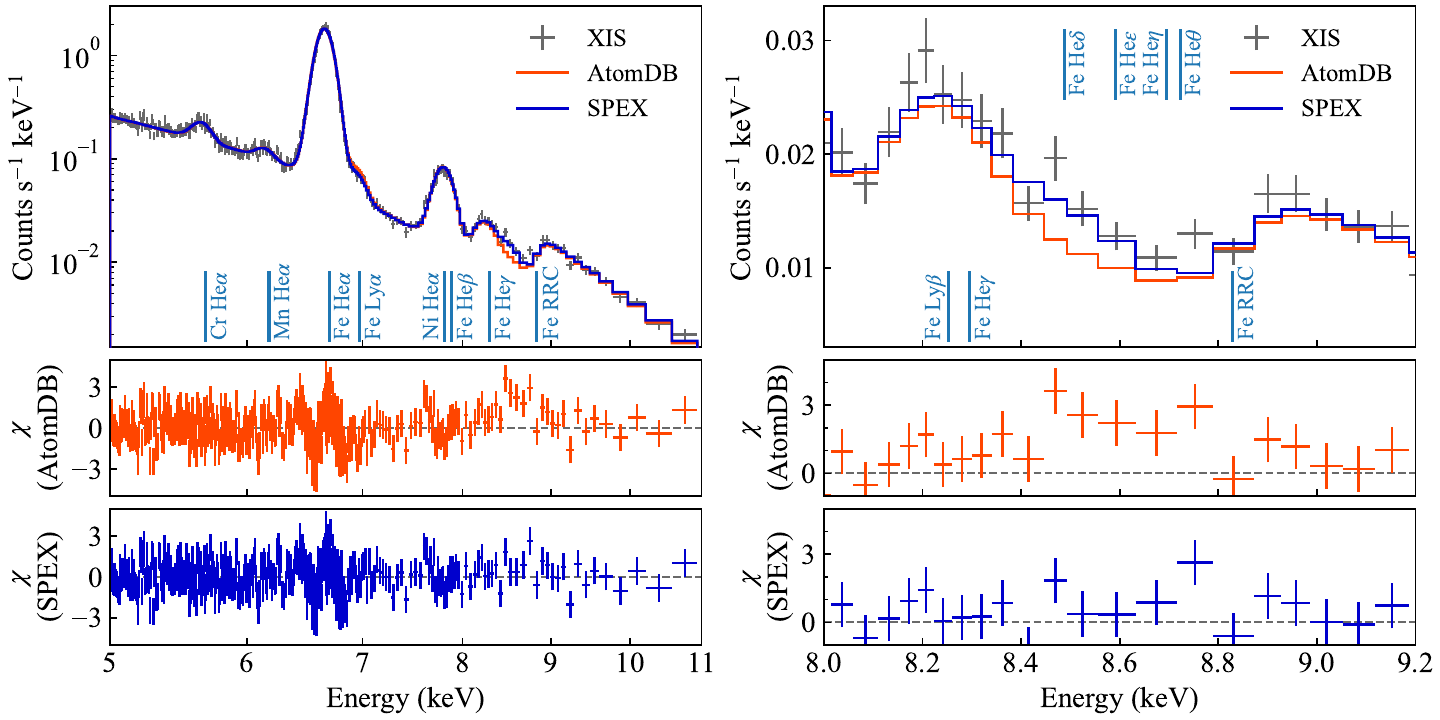}
\end{center}
\caption{X-ray spectrum of W\,49\,B with XIS\,0 and XIS\,3 combined in 5.0-11.0~keV (left) and its close-up view in 8.0--9.2~keV. The top panels show the observed spectrum and the best-fit model with the gain adjustment applied. The middle and bottom panels show the residuals of the data from the best-fit model, normalized by the standard error, for the fits using AtomDB and SPEX, respectively.
}
\label{fig:specfit}
\end{figure*}

First, as done in \citet{Suzuki2024}, we used Xspec version 12.13.1 with AtomDB version 3.0.9 to fit the observed spectrum, i.e., \texttt{vvrnei} for recombining plasma emission and \texttt{tbabs} for interstellar absorption. The observed spectrum and best-fit model are shown in Figure~\ref{fig:specfit}. There are two notable residual features (the middle panels). One is M-shaped feature around the Fe He$\alpha$ line and the other is a broad bump-like feature in 8--9~keV. These are the structures that \citep{Suzuki2024} argued to be the neutral Fe\emissiontype{I} K$\alpha$ line and Fe\emissiontype{XXV} CX emission, respectively. 

Next, we analyzed the same spectrum using an alternative spectral code, SPEX, with its Atomic Code and Tables version 3.08.01. We employed essentially the same emission model, using \texttt{neij} as the recombining plasma emission and \texttt{hot} as the interstellar absorption. The latter model is capable of modeling absorption by a CIE plasma, but we fixed the absorber temperature to the minimum value of 1~meV so that it acts as absorption by neutral materials. The best-fit model with SPEX is also shown in Figure~\ref{fig:specfit}. As shown in the bottom panels, especially in the magnified subfigure on the right, it is clear that the bump-like residuals in 8--9 keV are much less noticeable with SPEX. 

The best-fit parameters with AtomDB and SPEX are summarized in Table~\ref{tab:bestfit}. Between the two atomic codes, the parameters that describe thermal properties of the recombining plasma, i.e., $kT_{\rm init}$, $kT_{\rm e}$, and $\tau$, agree with each other within the statistical errors. The elemental abundances and $EM$ show apparently large discrepancies. However, when these are multiplied to infer the ion density, it showed better agreement with a typical discrepancy of $\approx 20$\%.

The gain adjustment affected the derived thermal parameters ($kT_{\rm init}$, $kT_{\rm e}$, and $\tau$). For instance, the recombination timescale $\tau$ showed more than one-order-of-magnitude difference, which can be regarded as a measure of potential systematic uncertainty in these parameters with the spectroscopy using X-ray CCDs.

\begin{table*}[htp]
\caption{Best-fit parameters for a recombining NEI plasma model.}
\begin{center}
\begin{tabular}{llcccccc}
\toprule
Parameter & Spectral code & \multicolumn{2}{c}{AtomDB v3.0.9} & \multicolumn{2}{c}{SPEX v3.08.01} \\
 & Gain adjustment & No & Yes & No & Yes \\
\midrule
$EM$ & ($10^{59}$~cm$^{-3}$) & $3.31_{-0.15}^{+0.15}$ & $3.07_{-0.13}^{+0.13}$ & $2.30_{-0.10}^{+0.10}$ & $2.27_{-0.09}^{+0.09}$  \\
$kT_{\rm init}$ & (keV) & $3.16_{-0.11}^{+0.13}$ & $2.96_{-0.04}^{+0.04}$ & $3.09_{-0.11}^{+0.12}$ & $2.84_{-0.04}^{+0.04}$ \\
$kT_{\rm e}$ & (keV) & $1.40_{-0.02}^{+0.03}$ & $1.44_{-0.02}^{+0.03}$ & $1.41_{-0.03}^{+0.03}$ & $1.45_{-0.03}^{+0.03}$ \\
$\tau$ & ($10^{10}$~cm$^{-3}$~s) & $10_{-3}^{+4}$ & $< 0.7$ & $13_{-3}^{+3}$ & $< 1.8$ \\
$Z_{\rm Cr}$ & (solar) & $5.33_{-0.54}^{+0.57}$ & $4.88_{-0.46}^{+0.48}$ & $8.40_{-0.90}^{+0.95}$ & $7.39_{-0.79}^{+0.83}$ \\
$Z_{\rm Mn}$ & (solar) & $4.62_{-0.79}^{+0.81}$ & $3.86_{-0.68}^{+0.65}$ & $7.15_{-1.33}^{+1.39}$ & $5.89_{-1.15}^{+1.19}$ \\
$Z_{\rm Fe}$ & (solar) & $3.84_{-0.16}^{+0.18}$ & $3.49_{-0.09}^{+0.10}$ & $6.60_{-0.30}^{+0.33}$ & $5.67_{-0.22}^{+0.25}$ \\
$Z_{\rm Ni}$ & (solar) & $4.38_{-0.65}^{+0.70}$ & $2.98_{-0.46}^{+0.48}$ & $5.73_{-0.88}^{+0.94}$ & $4.15_{-0.69}^{+0.73}$ \\
\midrule
$\chi^2$/dof & & 604.1/368 & 575.5/368 & 528.8/368 & 504.0/368 \\
\bottomrule
\end{tabular}
\end{center}
\label{tab:bestfit}
\end{table*}%

\subsection{Additional Fe K-shell lines}

We then searched for the previously claimed neutral Fe\emissiontype{I} K$\alpha$ at 6.4~keV and Fe\emissiontype{XXV} CX emission feature in 8--9~keV. These were examined separately by adding a Gaussian on top of the baseline best-fit models obtained in \S\ref{sec:fit} with all the spectral parameters fixed except for $EM$ (normalization). For Fe\emissiontype{I} K$\alpha$, a narrow Gaussian with a fixed 1$\sigma$ width of 0.0~keV was used, while for Fe\emissiontype{XXV} CX, a broad Gaussian with a fixed 1$\sigma$ width of 0.2~keV was used, following \citet{Suzuki2024}. Unlike the previous study, we allowed the centroids to be free for the both features. The line centroids, intrinsic (unabsorbed) flux and emission rates assuming the 11.3~kpc distance, and improvement of the fits measured as the change in the fit statistic are summarized in Table~\ref{tab:gauss}. 

The narrow line added near 6.4~keV consistently showed a higher centroid of 6.44--6.46~keV than that of neutral K$\alpha$ line at 6.40~keV regardless of the choice of the spectral code or gain adjustment, indicating a possibility that the feature rather originates from mildly ionized ions of Fe, although the discrepancy is statistically marginal. The higher centroid also means that the feature is closer to the bright Fe\emissiontype{XXV} He$\alpha$ line complex, indicating another possibility that the excess is caused by an imperfect calibration of the line-spread function far from its Gaussian core, although we cannot evaluate the magnitude of such an effect as it was out of the scope of the in-orbit calibration of the XIS RMF. The statistical significance of the added line, evaluated by dividing the measured rate by its lower statistical error converted to 1$\sigma$, ranges between 2--4$\sigma$. If the two results with the gain adjustment are compared to each other, there is no difference in the measured flux or the detection significance between the two spectral codes. Also, the measured flux is consistent with the previous report \citep{Suzuki2024} within the statistical errors. Therefore, we will not discuss this feature further in this paper. 

The broad line added in 8--9~keV showed variations in the measured line centroids and detection significances depending on the spectral code or gain adjustment. In particular, a clear contrast was found between the spectral codes; the detection significance was reduced from 6--7$\sigma$ (AtomDB) to 2--3$\sigma$ (SPEX), which is consistent with the trend in the spectral fit residuals (Figure~\ref{fig:specfit}). This result indicates a possibility that a significant fraction of the residual fluxes are actually due to the systematic uncertainty in the atomic codes. We will discuss this further in \S\ref{sec:cx}. 

\begin{table*}[htp]
\caption{Constraints on additional Gaussians.}
\begin{center}
\begin{tabular}{llcccccc}
\toprule
Parameter & Spectral code & \multicolumn{2}{c}{AtomDB v3.0.9} & \multicolumn{2}{c}{SPEX v3.08.01} \\
 & Gain adjustment & No & Yes & No & Yes \\
\midrule
\multicolumn{6}{c}{\dotfill Narrow line in 6.4--6.5~keV\dotfill} \\
Centroid & (keV) & $6.46_{-0.04}^{+0.03}$ & $6.44_{-0.04}^{+0.04}$ & $6.44_{-0.07}^{+0.05}$ & $6.45_{-0.04}^{+0.04}$ \\
Width at $1\sigma$ & (keV) & 0.0 (fixed) & 0.0 (fixed) & 0.0 (fixed) & 0.0 (fixed) \\
Flux & ($10^{-5}$ s$^{-1}$ cm$^{-2}$) & $1.11_{-0.50}^{+0.56}$ & $0.58_{-0.30}^{+0.32}$ & $0.52_{-0.38}^{+0.48}$ & $0.59_{-0.30}^{+0.37}$ \\
Rate & ($10^{41}$ s$^{-1}$) & $1.69_{-0.77}^{+0.86}$ & $0.88_{-0.46}^{+0.49}$ & $0.80_{-0.59}^{+0.72}$ & $0.89_{-0.46}^{+0.56}$ \\
\midrule
$\Delta \chi^2$ & & $-16.9$ & $-10.7$ & $-4.7$ & $-9.9$ \\
\midrule
\multicolumn{6}{c}{\dotfill Broad line in 8.4--8.8~keV\dotfill} \\
Centroid & (keV) & $8.47_{-0.09}^{+0.09}$ & $8.55_{-0.10}^{+0.10}$ & $8.60_{-0.28}^{+0.24}$ & $8.72_{-0.35}^{+0.24}$ \\
Width at $1\sigma$ & (keV) & 0.2 (fixed) & 0.2 (fixed) & 0.2 (fixed) & 0.2 (fixed) \\
Flux & ($10^{-5}$ s$^{-1}$ cm$^{-2}$) & $1.88_{-0.43}^{+0.42}$ & $1.37_{-0.38}^{+0.38}$ & $0.78_{-0.40}^{+0.40}$ & $0.52_{-0.39}^{+0.40}$ \\
Rate & ($10^{41}$ s$^{-1}$) & $2.86_{-0.66}^{+0.64}$ & $2.09_{-0.58}^{+0.58}$ & $1.18_{-0.61}^{+0.61}$ & $0.79_{-0.60}^{+0.60}$ \\
\midrule
$\Delta \chi^2$ & & $-56.5$ & $-35.6$ & $-10.1$ & $-4.8$ \\
\bottomrule
\end{tabular}
\end{center}
\label{tab:gauss}
\end{table*}%

\subsection{Abundance ratios} \label{sec:abratios}

\begin{figure*}[htbp]
\begin{center}
\includegraphics[width=8cm]{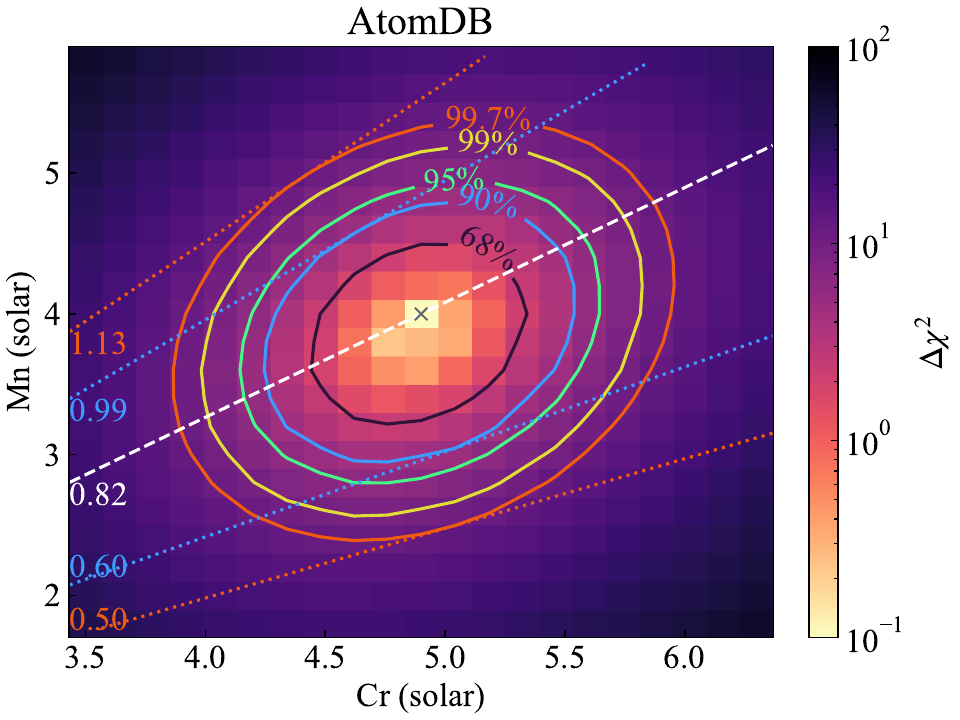}
\includegraphics[width=8cm]{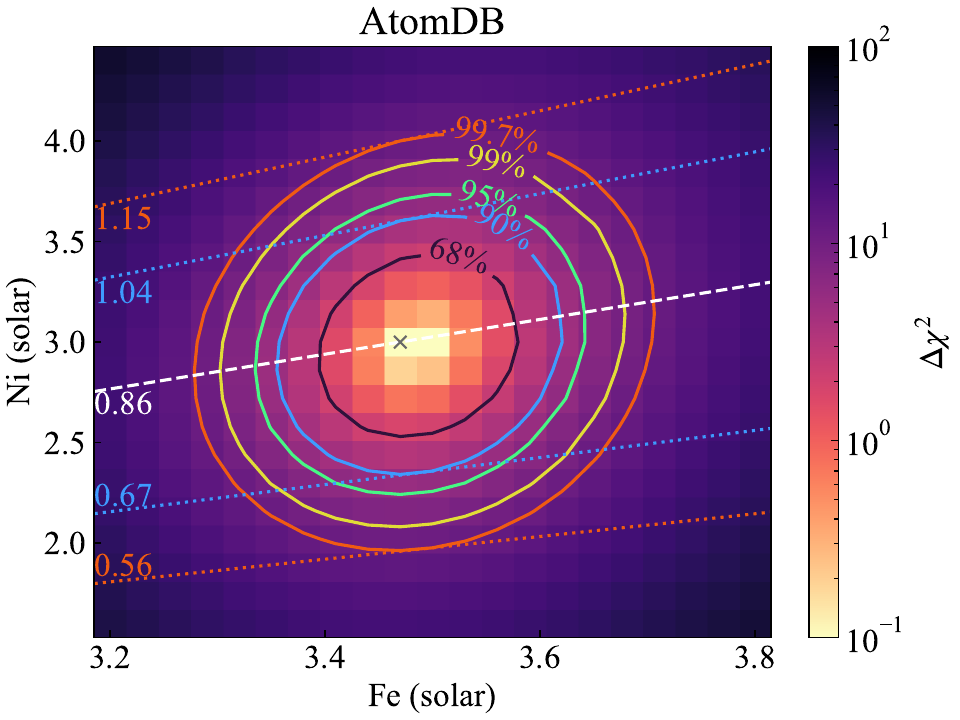}
\includegraphics[width=8cm]{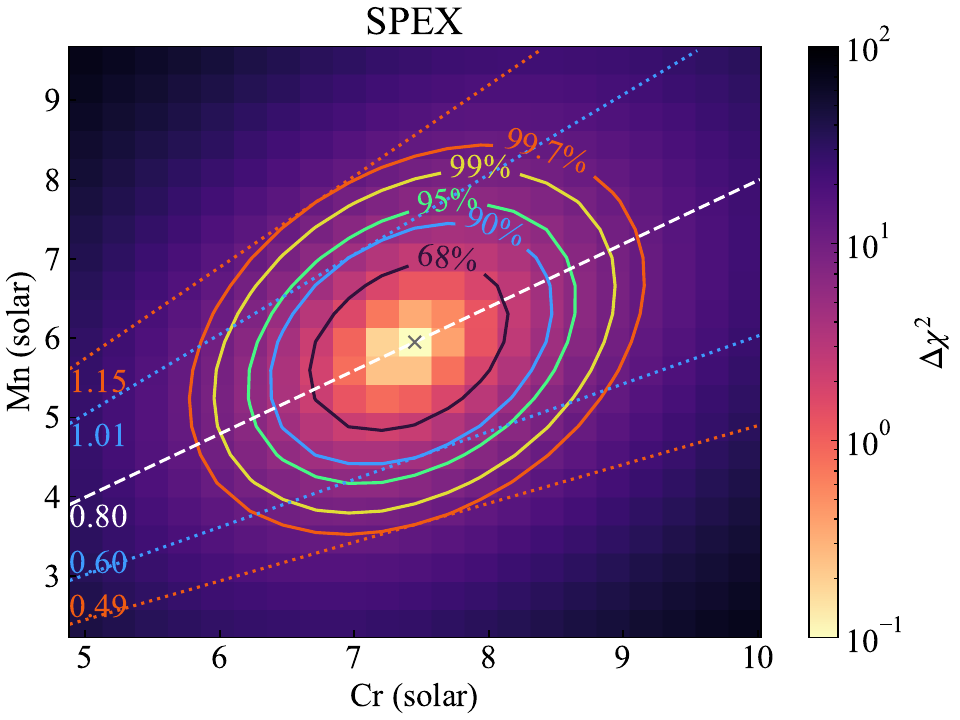}
\includegraphics[width=8cm]{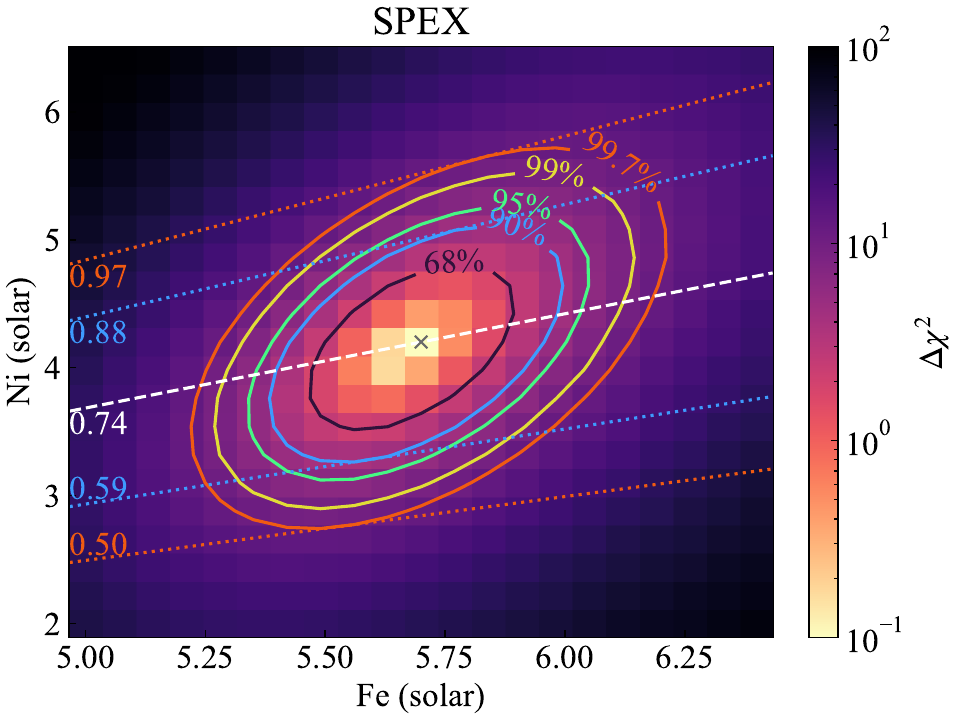}
\end{center}
\caption{Two dimensional distribution of the change in the fit statistic ($\Delta \chi^2$) for the Mn vs. Cr abundances (left panels) and the Ni vs. Fe abundances (right panels) for the fittings with the gain adjustment. Top panels are for the results with AtomDB and the bottom panels are for SPEX.
}
\label{fig:steps}
\end{figure*}

The abundance ratios can be derived from the best-fit values of $Z_X$, but its uncertainty may be larger than the one propagated from single-parameter errors due to possible correlations. For instance, at the CCD resolution, Fe He$\beta$ and Ni He$\alpha$ are blended, thus the abundances of these are likely correlated. We therefore evaluated two-parameter confidence ranges for each of $Z_{\rm Mn}$ vs. $Z_{\rm Cr}$ and $Z_{\rm Ni}$ vs. $Z_{\rm Fe}$ by sampling $\chi^2$ values in the parameter space using \texttt{steppar} in Xspec and \texttt{step} in SPEX. The results are shown in Figure~\ref{fig:steps}. The Mn/Cr abundance ratio agrees well between the two atomic codes. On the other hand, the Ni/Fe abundance ratio shows some discrepancy, although statistically marginal. 

\section{Discussion} \label{sec:disc}

\subsection{Charge exchange emission} \label{sec:cx}

\begin{figure*}[htbp]
\begin{center}
\includegraphics[width=17cm]{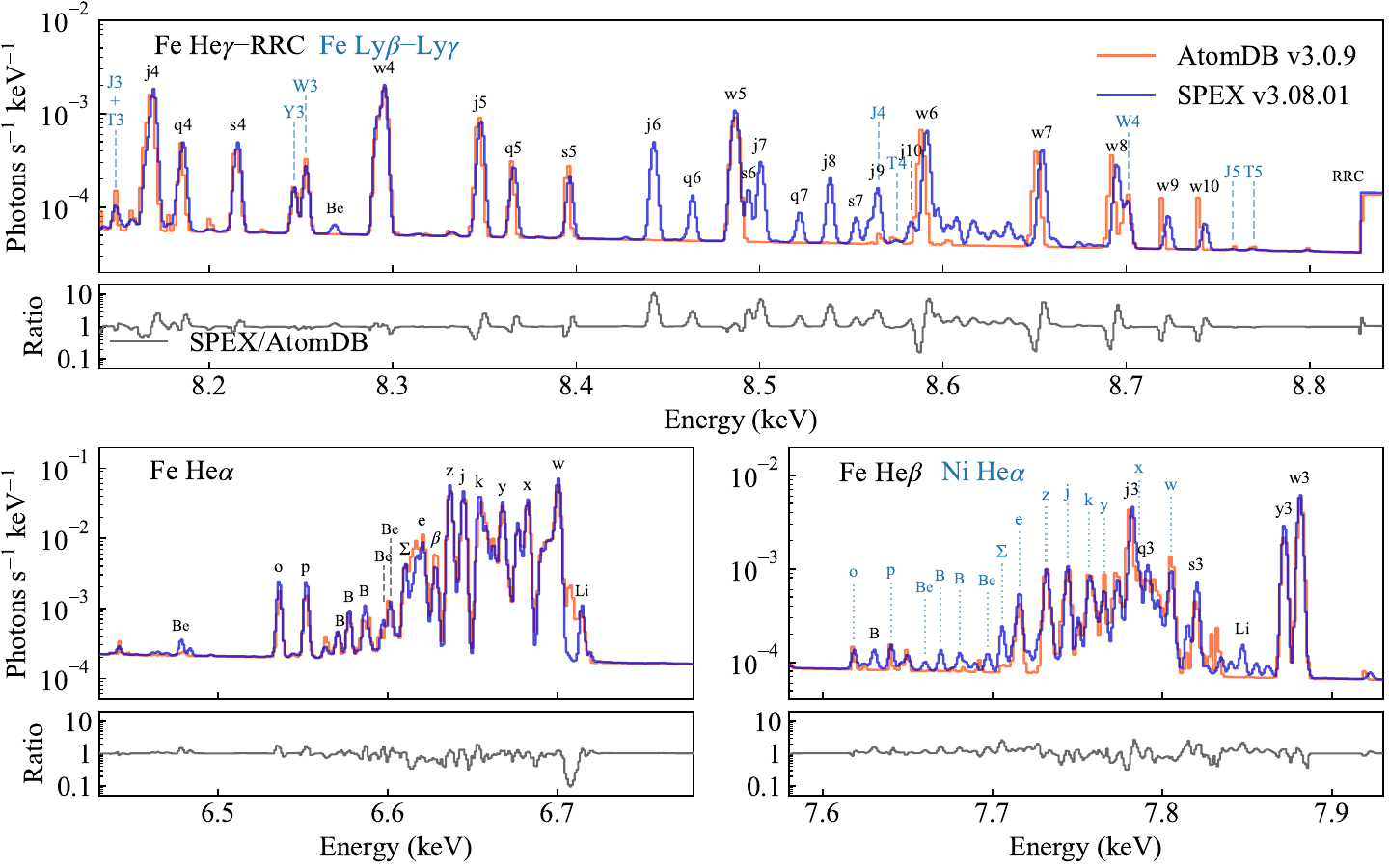}
\end{center}
\caption{Spectral model comparison in Fe He$\gamma$--RRC (top) as well as those in Fe He$\alpha$ (bottom left), Fe He$\beta$ (bottom right). In each subfigure, the top panel shows model spectra calculated with AtomDB v3.0.9 (red) and SPEX v3.08.01 (blue), while the bottom panel shows the ratio between the two. The key letters for line emission are mostly the same as those in Table~11 of \citet{Hitomi2018} and references therein. In particular, the 1s--$n$p resonance (intercombination) transitions for He-like ions are denoted as w$n$ (y$n$), and those for H-like ions are denoted as W$n$ (Y$n$), i.e., w4 is He$\gamma$ and W3 is Ly$\beta$. Minor DRs from B-like, Be-like, and Li-like ions, which are denoted as B, Be, and Li, respectively.
}
\label{fig:models}
\end{figure*}

The charge exchange emission reported by \citet{Suzuki2024} had the flux of $(1.96 \pm 0.23) \times10^{-5}$~s$^{-1}$~cm$^{-2}$, which is consistent with our result using AtomDB (Table~\ref{tab:gauss}). In our analysis, we found that the flux was reduced by nearly a factor of three when SPEX was used. Combined with the gain adjustment, our result showed a four times smaller flux than \citet{Suzuki2024}. To understand the discrepancy, we compare the model spectra of the recombining plasma component in the Fe He$\gamma$--RRC (+ Fe Ly$\beta$--Ly$\gamma$) band as well as those in Fe He$\alpha$ and Fe He$\beta$ (+ Ni He$\alpha$) in Figure~\ref{fig:models}. 

The model spectra of the two spectral codes agree well for most of bright lines such as Fe\emissiontype{XXV} He$\alpha$ triplet (the resonance line w, inter-combination lines x and y, and forbidden line z) and its dielectronic recombination (DR) satellite lines (denoted as j, k, o, and p). The agreement continues up to $n=5$ transitions ($n$ denotes the principal quantum number of the upper level of the transition), both for the resonance line (w$n$) and satellite lines (j$n$ for the DR satellite and q$n$ for the inner-shell excitation satellite, for instance). For the satellites lines, however, no further lines are included passed $n=5$ in AtomDB. Therefore, more than half of the possible CX flux reported by \citet{Suzuki2024} must be attributed to these satellite lines currently missing from AtomDB. 

\begin{figure*}[htbp]
\begin{center}
\includegraphics[width=17cm]{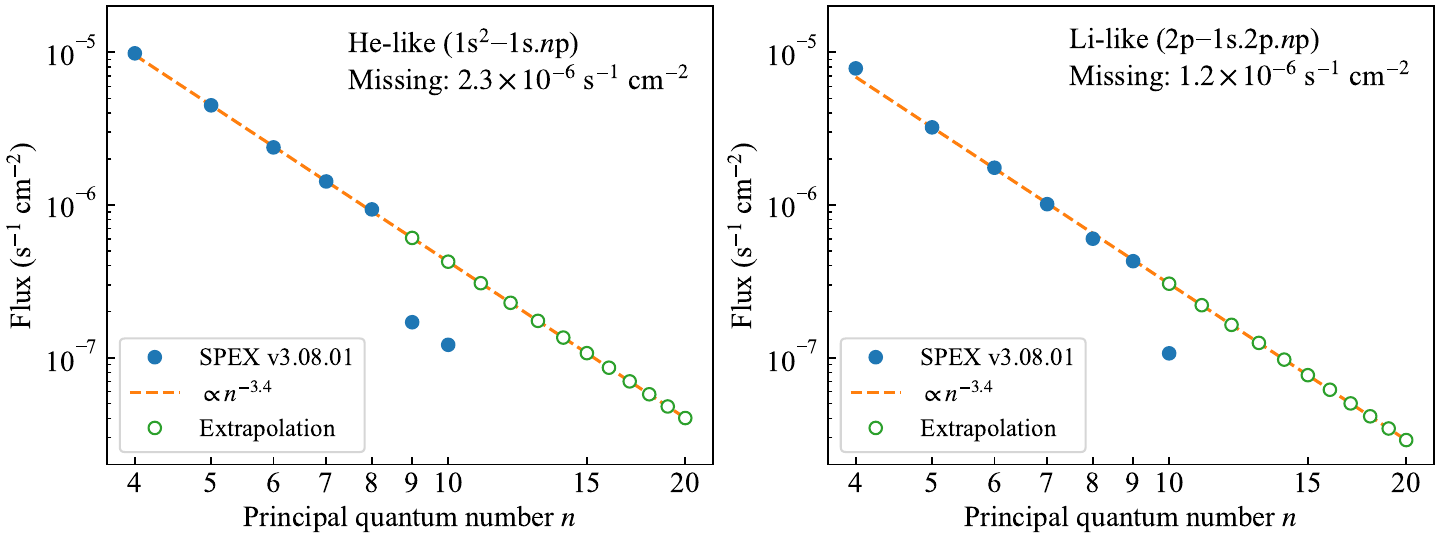}
\end{center}
\caption{Distributions of line fluxes calculated with SPEX over the upper-level principal quantum number for He-like ions (left) and Li-like ions (right).
}
\label{fig:resflux}
\end{figure*}

The missing lines in AtomDB lead us to a possibility that SPEX is also missing some lines. This is indeed the case; for instance, there are no lines above $n=10$ in both spectral codes as seen in Figure~\ref{fig:models}. Furthermore, the resonance line fluxes of $n=9$ and $n=10$ are significantly lower than those expected from the trend extrapolated from lower-$n$ lines. Similarly, the DR flux suddenly drops at $n=10$. These are more visually manifested in Figure~\ref{fig:resflux}. The model flux up to $n=8$ for the He-like lines (1s$^2$--1s.$n$p resonance and inter-combination lines) and $n=9$ for the Li-like lines (2p--1s.2p.$n$p satellite lines) shows a power-law dependence on $n$ with an index of $\approx -3.4$. The sudden drop of the flux is due to a known limitation on the current spectral calculation in SPEX; for instance, energy levels of He-like Fe ions are included up to $n=16$, but electron collisional excitation and radiative recombination are calculated only up to $n=10$ and $n=8$, respectively. By comparing the extrapolation of this trend to higher-$n$ lines, we estimate the missing flux in SPEX to be $2.3 \times 10^{-6}$~s$^{-1}$~cm$^{-2}$ and $1.2 \times 10^{-6}$~s$^{-1}$~cm$^{-2}$ for the He-like and Li-like lines, respectively. These add up to $\approx 45$\% of the flux measured with the added broad Gaussian in the case of the SPEX fit with no gain adjustment, further reducing the significance to $\approx 1.8\sigma$. The effect is even larger for the SPEX fit with the gain adjustment; the missing flux accounts for $\approx 67$\% of the measured Gaussian flux, corresponding to the reduced significance of only $\approx 0.7\sigma$. In this case, the remaining flux is smaller than 9\% of the originally claimed excess by \citet{Suzuki2024}.

Our analysis used observations in 2009 only, and adding XIS\,3 data from those in 2015 would reduce the statistical errors and improve the significance by a factor of 1.7 (\S\ref{sec:obs}). We note that, even if this factor is taken into account, the detection significance of the possible CX feature would barely reach the $3\sigma$ level for the case with no gain adjustment. This is even reduced if we adopt the result with the gain adjustment. We note that, there are other missing lines, for instance, DR satellites of $n \ge 3$ from Be-like ions, which may explain the remaining flux.

Considering all these factors, we conclude that the previously reported excess in 8--9~keV is fully attributable to the systematic error in the spectral codes, and therefore, there is no observational evidence for the CX emission in the Suzaku spectrum of W\,49\,B. This conclusion, however, does not eliminate a case where the CX emission exists at a much lower flux level. To examine such a possibility, a high-resolution spectroscopy is required, such as using the X-ray Imaging and Spectroscopy Mission (XRISM: \citealt{Tashiro2022}).  

\begin{figure*}[htbp]
\begin{center}
\includegraphics[width=17cm]{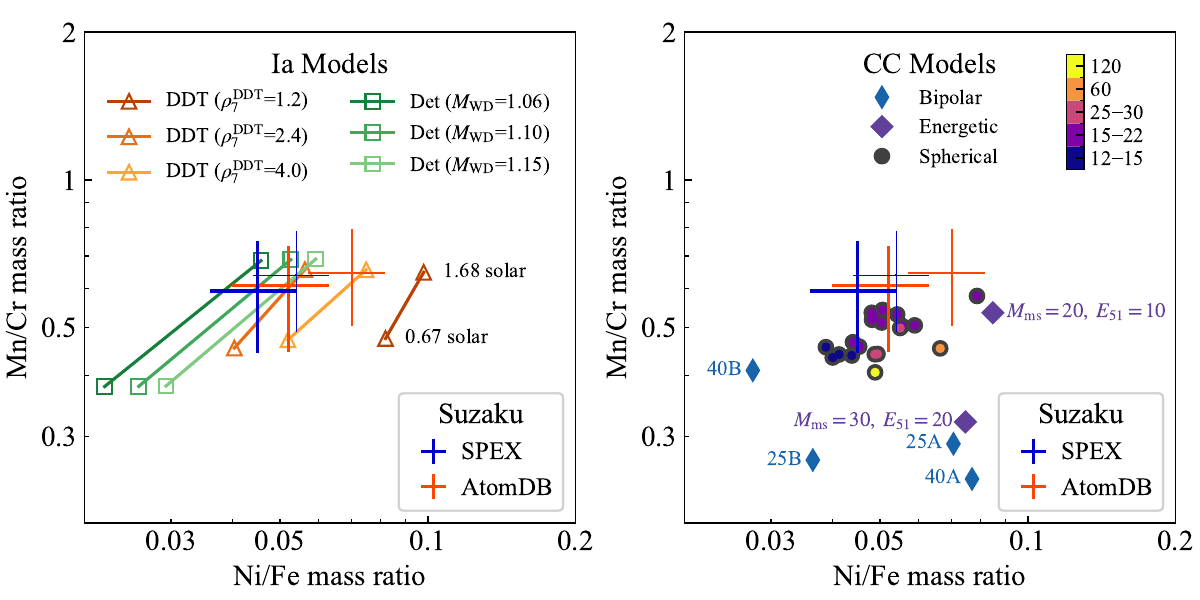}
\end{center}
\caption{The Mn/Cr vs. Ni/Fe ejecta mass ratios of W\,49\,B compared with nucleosynthesis models for Ia (left) and CC (right) explosions. The thick and thin crosses show the best-fit results with and without the gain adjustment, respectively, to indicate the magnitude of the systematic error due to the gain calibration uncertainty. The results with SPEX and AtomDB are shown separately. The Ia models for Chandrasekhar-mass delayed detonation (DDT) cases and sub-Chandrasekhar-mass pure detonation (Det) cases are from one-dimensional calculations by \citet{Bravo2019}. The CC models are from \citet{Maeda2003} for bipolar explosions,  \citet{Nomoto2006} for energetic explosions, and \citet{Sukhbold2016} for spherical explosions. The progenitor mass range for spherical explosions is indicated by the marker color as coded in the color bar in the right panel.
}
\label{fig:massratios}
\end{figure*}

\subsection{Fe-group ejecta mass ratios and progenitor} \label{sec:prog}

The ejecta mass ratios of the Fe-group elements, the Ni/Fe and Mn/Cr ratios, are also revised using the new constraints obtained in our analysis (\S\ref{sec:abratios}). Figure~\ref{fig:massratios} compares our measurements to expectations from various SN nucleosynthesis models. We find that the systematic difference due to the choices on the spectral code or the gain adjustment is larger for the Ni/Fe ratio than for the Mn/Cr ratio. This is reasonable considering the spectral blending of the Ni He$\alpha$ and Fe He$\beta$ complexes. 

For the Ia models (the left panel of Figure~\ref{fig:massratios}), we quote one-dimensional calculations by \citet{Bravo2019} with the metallicities close to the solar value. The calculations include results for the Chandrasekhar-mass delayed detonation (DDT) models (triangles) and the sub-Chandrasekhar-mass pure detonation (Det) models (squares). The DDT models are shown for three different values of the deflagration-to-detonation transition density ($\rho_7^{\rm DDT}$ in units of $10^7$~cm$^{-3}$). The Det models are for three different white dwarf masses ($M_{\rm WD}$ in units of $M_{\odot}$). There are overlaps between the measurements and the theoretical expectations from both the DDT and Det models, providing a support for the Ia SN origin of W\,49\,B, as previously claimed using abundance pattern of other elements \citep[e.g.,][]{Zhou2018}. One exception among the models quoted here would be the DDT model with $\rho_7^{\rm DDT} \sim 1$. A smaller transition density means a higher contribution from the deflagration, resulting in a higher Ni/Fe mass ratio. Thus, the DDT model with $\rho_7^{\rm DDT} \lesssim 1$ is likely rejected as the origin of this remnant. The DDT model with a significantly higher central density of the exploding white dwarf (e.g., as required for the Ia SNR 3C\,397: \citealt{Ohshiro2021}) than a typical value (as adopted in \citealt{Bravo2019}; Figure~\ref{fig:massratios}) is likely rejected, too, as such a model predicts a higher Ni/Fe mass ratio \citep{Leung2018}. 
The Det models predict systematically lower Ni/Fe ratios than the DDT models, but the difference is comparable to the systematic uncertainty in the spectroscopy. Considering the more complete emission line list available in SPEX, one may consider that, at present, the Ni/Fe ratio measured with SPEX would be more reliable. If this is indeed the case, the Det models would slightly be favored over the DDT models. We note that the previously claimed discrepancy in the Mn/Cr mass ratio from the models \citep{Sato2020} no longer exists in our measurements ($\approx 0.6$), which is nearly two times smaller than the previous measurement by \citet{Zhou2018}. The authors used AtomDB version 3.0.7, but we have confirmed that re-analyzing the Suzaku spectrum using this particular version of the spectral code does not solve the large discrepancy. Therefore, we speculate that the discrepancy is rather caused by the difference in the data quality, such as the energy resolution. 

For the CC models (the right panel of Figure~\ref{fig:massratios}), we compare three different groups: \citet{Maeda2003} for bipolar explosions (narrow diamonds, two variations for each of $40~M_{\odot}$ and $25~M_{\odot}$ progenitors),  \citet{Nomoto2006} for energetic explosions (wide diamonds, with different values of the main-sequence progenitor mass $M_{\rm ms}$ in units of $M_{\odot}$ and of the kinetic explosion energy $E_{51}$ in units of $10^{51}$~erg), and \citet{Sukhbold2016} for spherical explosions (circles, with various progenitor masses as indicated with the color bar). With our revised constraints, especially the reduced Mn/Cr ratio, now we find that there are overlap with the CC models, too. The high Mn/Cr of $\sim 0.6$ and low Ni/Fe ratios of $\sim 0.05$ are more in line with the spherical explosion models than the bipolar or energetic explosion models, and if a spherical CC is the origin, the progenitor mass is likely small ($\lesssim 20$~$M_{\odot}$). One of the energetic models with $M_{\rm ms} = 20$ and $E_{51}=10$ is close to the measurement with AtomDB without the gain adjustment, which showed the highest Ni/Fe mass ratio among the four measurements. However, the Ni/Fe ratio is likely more reliably determined with SPEX, which currently has a more complete set of Fe DR satellite lines. This indicates that the Ni/Fe ratio is likely $\lesssim 0.06$. For the other energetic or bipolar CC models, a systematic discrepancy from the measurements is found in their underpredicted Mn/Cr ratios. We note that, these model calculations \citep{Maeda2003, Nomoto2006} did not include the neutrino process, which is known to contribute to the Mn yield \citep{Sato2023}. A similar underprediction, but likely at a smaller magnitude, probably exists in the spherical CC models because of the ignorance of the neutrino wind in the nucleosynthesis, which would also contribute to the Mn yield via the neutrino-proton process \citep{Sukhbold2016}. There is a difference in the progenitor star models, too, which may also be reflected into the mass yields. Therefore, we caution that the apparent discrepancies in the Mn/Cr ratio between the explosion types and subtypes are possibly due to the differences in the physical processes considered in the nucleosynthesis calculations rather than the explosions mechanisms themselves. Another factor to consider regarding the model systematics is the fact that physical parameters of explosion models are not necessarily fully explored. For instance, a significant variation in theoretical mass yields was found in the parameter study of the jet-driven SNe by \citet{Leung2023}. In addition, effects such as the initial metallicity and stellar rotation of the progenitors would 
alter the neutron excess in the stellar interior and alter the Mn and Ni yields \citep[e.g.,][]{Limongi2018}.

Based on the discussions above, we conclude that currently both Ia and spherical CC models with a progenitor mass of $\lesssim 20~M_{\odot}$ can explain the observed Fe-group ejecta mass ratios, while the other CC models are not completely ruled out considering possible systematics not only in the spectroscopy but also in the nucleosynthesis calculations. Further investigations require progresses in both the theoretical and observational fields. 

\subsection{Origin of recombining plasma}

Finally, we briefly address the implications of our results to the origin of the recombining plasma in W\,49\,B. The lack of a strong photo-ionization source and the insignificance of the additional collisional ionization by shock-accelerated protons \citep{Sawada2024} narrow the viable mechanisms to the rapid electron cooling. There are two cooling scenarios that could have produced the recombining plasma: the adiabatic cooling due to the interaction with the dense CSM formed by the progenitor's stellar winds \citep{Itoh1989, Yamaguchi2018} and the conduction cooling due to collision with cold clouds \citep{Kawasaki2002, Sano2021}. The CX emission would be direct evidence for a collision of the hot ejecta with cold clouds. However, we have shown that the previously claimed detection of the CX emission was spurious feature originating from the incompleteness of atomic codes. Therefore, currently there is no strong support for the conduction cooling model in this regard.

The progenitor constraint would be an important test for the adiabatic cooling scenario as it is closely related to the circumstellar environment. Our revised ejecta mass ratios are consistent with the theoretical expectations for spherical CC explosion models with a progenitor mass of $\lesssim 20~M_{\odot}$. Such CC explosions may occur inside a dense CSM shell formed by a red supergiant star. Although the CSM model in the original rarefaction scenario by \citet{Itoh1989} was Type II-L like, a dense CSM also exists in more popular Type II-P SNe \citep[e.g.,][]{Yaron2017}, and there are accumulating clues that such an environment is common \citep{Forster2018, Bruch2023}. Another possibility would be Type IIn SNe \citep{Smith2017}. Indeed, \citet{Moriya2012} demonstrated that a CSM associated with a Type IIn SN is dense/massive enough at least to set up the high ionization state of shock-heated plasma. 

The Type Ia origin is also a strong candidate especially if the abundance measurements on other elements are combined \citep[e.g.,][]{Sato2024}. The rarefaction scenario would require a dense/massive CSM similar to the CC cases discussed above. It has been observed that some of Type Ia SNe occur inside a dense CSM environment \citep{Hamuy2003, Dilday2012, Jerkstrand2020}, and such an environment may be responsible for overluminous Ia SNe \citep{Maeda2023}. \citet{Court2024} simulated hydrodynamics and NEI for shocked ejecta of Ia SNe exploded inside a planetary nebula, and found that the high luminosity and energy centroid of Fe He$\alpha$ of W\,49\,B can be explained if the explosion occurred with a short delay ($\sim 1$~kyr) after the formation of the planetary nebula. Despite being rare, such an Ia SN with a dense CSM may be the origin of W\,49\,B, and it may also explain the fact that this remnant is the only recombining SNR with a possible Ia origin other than the recently claimed recombining plasma in Sgr~A~East \citep{XRISM2024_SgrAEast}.

\section{Summary} \label{sec:summary}

In this study, we have revisited the Suzaku Fe-K band spectrum of the Galactic SNR W\,49\,B. We performed spectral fittings using two representative spectral codes for collisional plasma, AtomDB and SPEX, and different treatments on the gain calibration adjustment. Between the results of the two codes, a large discrepancy is found in 8--9 keV. The comparison and detailed examination of the spectral codes lead us to a conclusion that the previously reported Fe\emissiontype{XXV} CX emission is fully attributable to missing fluxes in the spectral code, mainly of Fe\emissiontype{XXV} high-shell DR satellite lines. We further examined the ejecta mass ratios of the Fe-group elements, Mn/Cr and Ni/Fe. In our results, the Mn/Cr ratio consistently shows lower values than in the literatures, while the Ni/Fe ratio varied depending on the choice of spectral code or the gain adjustment. These ratios are consistent with Ia explosion models with a typical central density of an exploding white dwarf and spherical CC explosion models with a progenitor mass of $\lesssim 20~M_{\odot}$, while the other CC models cannot be completely ruled out if potentially large systematics in the nucleosynthesis models are considered. Our study consistently demonstrates the importance of the accurate understanding of the spectral codes even for the moderate-resolution spectra from X-ray CCDs. For further investigations, a high-resolution X-ray spectroscopy with XRISM is highly anticipated.

\begin{ack}
We are grateful Dr.~A.\,A.~Raasen and Prof.~Jelle~Kaastra for the early discussions in 2014 on the discrepancy between the Suzaku data and spectral models at high-shell transition Fe lines of a recombining plasma, which eventually led us to the understanding presented in this paper. We are also grateful Dr.~Liyi~Gu and Dr.~Adam~Foster for the discussions on the current implementations and limitations of the spectral codes, SPEX and AtomDB, respectively. 
This work was supported by the RIKEN Pioneering Project Evolution of Matter in the Universe and Rikkyo University Special Fund for Research (M.S.), JSPS KAKENHI Grant Numbers JP23K13128 (T.S.), JP24H01810, and JP24KK0070, and The Kyoto University Foundation (K.M.).
\end{ack}

\bibliographystyle{aa} 
\bibliography{ms} 

\end{document}